\documentclass{PoS}

\title{Results of Nucleon Resonance Extraction via Dynamical Coupled-Channels Analysis from Collaboration@EBAC}

\ShortTitle{Dynamical Coupled-Channels Analysis from Collaboration@EBAC}

\author{\speaker{Hiroyuki Kamano}\\%
Research Center for Nuclear Physics (RCNP), Osaka University, Ibaraki, Osaka 567-0047, Japan\\
       E-mail: \email{kamano@rcnp.osaka-u.ac.jp}}

\abstract{
We review a global analysis of meson production reactions off the nucleons by 
a collaboration at Excited Baryon Analysis Center of Jefferson Lab.
The analysis is pursued with a dynamical coupled-channels approach, within which the dynamics
of multi-channel reaction processes are taken into account in a fully consistent way with
the two-body as well as three-body unitarity of the S-matrix.
With this approach, new features of nucleon excitations are revealed as resonant particles
originating from the non-trivial multi-channel reaction dynamics, 
which cannot be addressed by static hadron models
where the nucleon excitations are treated as stable particles.
}

\FullConference{Sixth International Conference on Quarks and Nuclear Physics\\
                April 16-20, 2012\\
                Ecole Polytechnique, Palaiseau,  Paris}

\begin{document}

\section{Introduction}
\label{sec1}

An understanding of the spectrum and internal structure of the excited nucleons 
($N^\ast$) remains to be a fundamental challenge in the hadron physics. 
In fact, since the late 90s, a huge amount of high precision data of 
meson photoproduction reactions off the nucleons has been continuously reported 
from electron- and photon-beam facilities such as
ELSA, GRAAL, JLab, MAMI and SPring-8~\cite{bl04}.
This experimental effort enables one to make a quantitative determination of 
the $N^\ast$ properties directly from the data.
Theoretical researches are also being made actively on the basis of
the Lattice QCD~(see e.g., Refs.~\cite{edw11,de12,hwl12,ale11}) and hadron models such as 
the Dyson-Schwinger equations~(see e.g., Refs.~\cite{hll11,djw12}).

\begin{figure}[b]
\begin{center}
\includegraphics[width=.4\textwidth,clip]{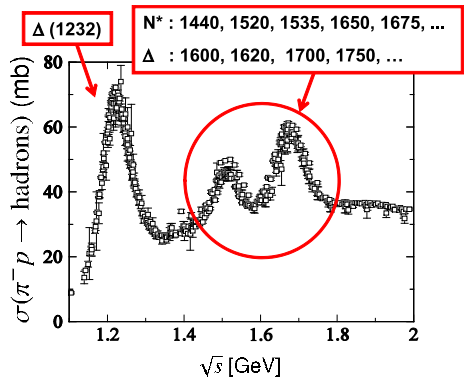}
\caption{Peaks in the $\pi^- p$ reaction total cross sections.
The first peak at $\sqrt{s} \sim 1.2$ GeV is known to be produced by
$\Delta(1232)$, the first $P_{33}$ nucleon resonance,
while the second and third peaks contain more than 10 $N^\ast$ states,
which indicates that those are highly overlapping with each other in energy.
\label{fig:totcs}
}
\end{center}
\end{figure}

Besides its important role in understanding QCD of the nonperturbative domain, 
the $N^\ast$ spectroscopy is an interesting subject for studying resonance phenomena. 
Because of their strong couplings to (multi-channel) meson-baryon continuum 
states, the $N^\ast$s appear as very broad and highly overlapping resonances 
in the $\pi N$ and $\gamma N$ reaction cross sections.
In fact, decay widths of the $N^\ast$s are about a few hundred MeV in average, 
and, for example, the second and third peaks in the $\pi^- p$ total cross 
section turn out to contain more than 10 $N^\ast$ states as shown in 
Fig.~\ref{fig:totcs}. 
This indicates that a peak in the cross sections does not 
necessarily mean the existence of an isolated resonance at the peak.
This situation is quite different from resonances of other systems such as 
heavy-quark hadrons and nuclei. 
In those systems, each resonance usually appears as a clear and well-separated 
peak in the cross sections.
The broad and overlapping nature of the $N^\ast$s make themselves into a unique 
laboratory to study the resonance physics in the multi-channel reaction processes.

In order to explore the nature of $N^\ast$ states, a collaboration was developed at 
the Excited Baryon Analysis Center (EBAC) of Jefferson Lab in the spring of 2006.
Main objectives of the collaboration are to determine the spectrum of the $N^\ast$ states and 
extract their form factors such as $N$-$N^\ast$ electromagnetic transition form factors 
through a global analysis of the meson production reactions with 
the $\pi N$, $\gamma N$, and $N (e,e')$ initial states in the resonance energy region.
Also, another important task is to provide reaction mechanism information 
which is necessary for making physical interpretations for the mass spectrum, 
quark-gluon substructures, and dynamical origins of the $N^\ast$ states.
In order to extract reliable information on the $N^\ast$ states from 
the reaction data, we need to develop a reliable model that describes all the 
relevant meson production reactions 
simultaneously in the wide energy and kinematic regions.
Because all the reactions are related with each other by the unitarity, 
the reaction model must take account of couplings among those reaction channels.
As such a reaction model, we employ a dynamical coupled-channels (DCC) approach developed 
in Ref.~\cite{msl07}, within which the couplings among relevant meson-baryon reaction channels 
including the three-body $\pi \pi N$ channel are fully taken into account, so that
the scattering amplitudes satisfy the two-body as well as three-body unitarity.
In this contribution, we review our 6-year effort on the global analysis of 
meson production reactions off the nucleons.

\section{Dynamical coupled-channels (DCC) approach to meson production reactions}
\label{sec2}

Our DCC approach is based on a multi-channel and 
multi-resonance model~\cite{msl07}, within which the partial wave amplitudes of 
$M(\vec p) + B(-\vec p) \to M'(\vec p') + B'(-\vec p')$
is obtained by solving the following coupled-channels integral equations:
\begin{eqnarray}
T_{M'B',MB}(p',p;E) &=& 
V_{M'B',MB}(p',p;E)
\nonumber\\
&&
+ \sum_{M''B''}\int_C dq q^2 V_{M'B',M''B''}(p',q;E) G_{M''B''}(q;E) T_{M''B'',MB}(q,p;E),
\label{eq:lseq}
\end{eqnarray}
where $MB,M'B',M''B''=\pi N, \eta N, \pi\Delta, \rho N, \sigma N, K\Lambda, K\Sigma,\cdots$.

\begin{figure}[t]
\begin{center}
\includegraphics[width=.75\textwidth,clip]{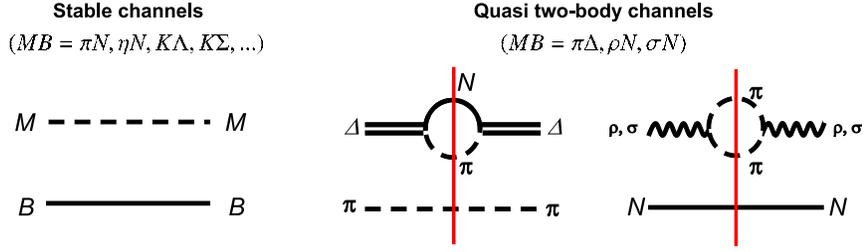}
\caption{
Meson-baryon Green functions $G_{M'B',MB}(p,p';E)$.
For the quasi-two-body Green functions, the three-body $\pi\pi N$ cut is produced at
vertical lines.
\label{fig:green}
}
\end{center}
\end{figure}

The meson-baryon Green functions are represented as
$G_{MB}(k,E)=1/[E-E_M(k)-E_B(k) + i\epsilon]$ for the stable $MB$ channels such as
$\pi N$ and $\eta N$, while $G_{MB}(k,E)=1/[E-E_M(k)-E_B(k) -\Sigma_{MB}(k,E)]$
for the quasi-two body $\pi\Delta$, $\rho N$, and $\sigma N$ channels
that can decay into the 3-body $\pi\pi N$.
The self energy
$\Sigma_{MB}(k,E)$ is calculated from a vertex function defining the decay of
the considered unstable particle in the presence of a spectator $\pi$ or $N$ 
with momentum $k$. For example, we have for the $\pi\Delta$ state,
\begin{equation}
\Sigma_{\pi\Delta}(k,E) =
\frac{m_\Delta}{E_\Delta(k)}
\int_{C'} q^2 dq 
\frac{ M_{\pi N}(q)}{[M^2_{\pi N}(q) + k^2]^{1/2}}
\frac{\left|f_{\Delta \to \pi N}(q)\right|^2}
{E-E_\pi(k) -[M^2_{\pi N}(q) + k^2]^{1/2} + i\epsilon},
\label{eq:self-pid}
\end{equation}
where $E_\alpha(k)=[m^2_\alpha + k^2]^{1/2}$ with $m_\alpha$ being
the mass of particle $\alpha$; $M_{\pi N}(q) =E_\pi(q)+E_N(q)$; $f_{\Delta \to \pi N}(q)$
defines the decay of the $\Delta \to \pi N$ in the rest frame of $\Delta$. 
The self-energies for the $\rho N$ and $\sigma N$ channels have similar forms.
With this appropriate consideration of the self-energies,
the three-body $\pi\pi N$ cut is correctly maintained in the resulting partial waves amplitudes.

\begin{figure}[t]
\begin{center}
\includegraphics[width=.8\textwidth,clip]{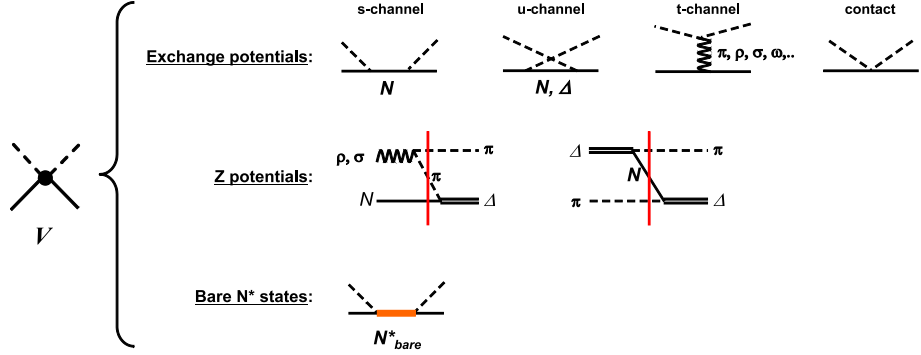}
\caption{
Transition potential $V_{M'B',MB}(p',p;E)$.
Vertical lines in the $Z$-potentials indicate that the three-body $\pi\pi N$ cuts 
are produced there.
\label{fig:pot}
}
\end{center}
\end{figure}

The $MB\to M'B'$ transition potential $V_{M'B',MB}(p',p;E)$ consists of three pieces
(a diagrammatic representation of the potential is shown in Fig.~\ref{fig:pot}):
\begin{equation}
V_{M'B',MB}(p',p;E) =
v_{M'B',MB}(p',p) + Z_{M'B',MB}(p',p;E) +
\sum_{N^\ast_i} \frac{\Gamma^\dag_{N^\ast_i,M'B'}(p')\Gamma_{N^\ast_i,MB}(p)}{E-m^0_{N^\ast_i}}.
\label{eq:pot}
\end{equation}
The first term $v_{M'B',MB}(p',p)$ is the so-called exchange potential, which
just contains ground-state hadrons of each spin-flavor multiplet.
The $v_{M'B',MB}(p',p)$ potential is derived from the effective Lagrangians 
using the unitary transformation method~\cite{sl96,sl09}.
It is noted that this method makes the exchange potential energy-independent and
uniquely defines its off-shell behavior.
The second term $Z_{M'B',MB}(p',p;E)$ is the so-called $Z$-potential, which 
appears due to eliminating the 3-body $\pi\pi N$ channel from 
the channel-indices of the scattering amplitudes
by using the Feshbach projection techniques. 
(See Appendix B of Ref.~\cite{msl07} for the details.)
In fact, the $Z$-potentials contain the three-body $\pi\pi N$ cut in the intermediate process.
Implementation of both the $Z$-potentials and the self-energies in the meson-baryon Green functions 
is necessary for maintaining the three-body unitarity in the amplitudes, and 
this makes our model quite unique among the existing models of meson production reactions.
The third term describes the propagation of the ``bare'' $N^\ast$ states, in which
$m^0_{N^\ast_i}$ and $\Gamma^\dag_{N^\ast_i,MB}(p)$ represent the bare mass of 
the $i$-th $N^\ast$ state and the bare $N^\ast_i \to MB$ decay vertex, respectively.

It is noted that within our model, the bare $N^\ast$ states are defined as the eigenstates 
of the Hamiltonian in which the couplings to the meson-baryon continuum states are turned off.
Therefore, by definition, our bare $N^\ast$ states can be related with 
the hadron states obtained from the static hadron models such as constituent quark models
and Dyson-Schwinger approaches.
Through the reaction processes, the bare $N^\ast$ states couple to the reaction channels
and then become resonance states.
Of course, there is another possibility that these two potentials generate resonance poles 
dynamically. 
Our approach can treat both mechanisms of resonance productions on the same footing.

\section{Physics highlights from the DCC analysis in 2006-2009}
\label{sec3}

During the developing stage in 2006-2009, we constructed a DCC model
including $\gamma N$, $\pi N$, $\eta N$, $\pi\Delta$, $\sigma N$, $\rho N$ channels.
Hadronic and electromagnetic parameters of the model are determined by 
analyzing $\pi N \to \pi N$~\cite{jlms07} and $\pi N \to \eta N$~\cite{djlss08}
up to $W=2$ GeV, and also $\gamma N \to \pi N$~\cite{jlmss08} and $N(e,e'\pi)N$~\cite{jklmss09}
up to $W=1.6$ GeV and $Q^2 = 1.5$ (GeV/c)$^2$. 
Then the model is applied to $\pi N\to \pi\pi N$~\cite{kjlms09} and
$\gamma N\to\pi\pi N$~\cite{kjlms09-2} to predict cross sections and examine consistency of 
the coupled-channels framework.
Also, making use of the analytic continuation method developed in Ref.~\cite{ssl09},
we have successfully extracted the $N^\ast$ mass spectrum~\cite{sjklms10,knls10} and 
the $N$-$N^\ast$ electromagnetic transition form factors~\cite{ssl10}
from the constructed model.

\begin{figure}[t]
\begin{center}
\includegraphics[width=0.5\textwidth,clip]{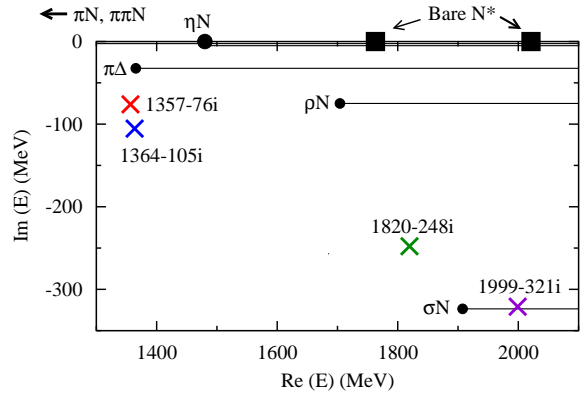}
\caption{
Resonance pole positions of $P_{11}$ nucleon resonances extracted from our early 
analysis~\cite{sjklms10} (indicated by crosses).
Filled squares in the real energy axis are the masses of bare $N^\ast$ states 
introduced in the DCC model; filled circles and lines starting from the circles are 
branch points and cuts of the reaction channels, respectively.
\label{fig:p11-pole}
}
\end{center}
\end{figure}

Figure~\ref{fig:p11-pole} shows the pole positions of the $P_{11}$ nucleon resonances
extracted from our early analysis.
There are three features of this figure, which clearly show the important role 
of the non-trivial multi-channel reaction dynamics in understanding the $N^\ast$ mass spectrum.
Firstly, two almost degenerate poles, $E=1357-76i$ and $1364-103i$ MeV, appear in the 
energy region where the Roper resonance is known to exist.
This result suggests that the Roper resonance is associated with these two resonance poles.
We have found that those can be regarded as the so-called pole and shadow-pole
with respect to the $\pi\Delta$ branch point. (See e.g. Ref.~\cite{eden-taylor} for the notion of
the pole and shadow-pole.)
The double-pole structure of the Roper resonance has been reported also by other analysis groups 
using completely different approaches~\cite{vpi85,said06,cmb90,julich09}.
Secondly, the number of the physical resonances is larger than that of 
the bare $N^\ast$ states. 
In fact, we have found three resonances for this partial wave\footnote{
We refer to the double poles at $E=1357-76i$ and $1364-103i$ MeV as the Roper resonance
collectively.}, 
whereas we introduced just two bare $N^\ast$ states.
One might think that some of the poles are just molecule-type resonances originating 
from purely meson-exchange processes.
However, at least within this analysis, such a possibility has been ruled out.
Therefore, this result shows that within the multichannel reaction processes, 
a naive one-to-one correspondence between the bare states (i.e., hadrons in static models) 
and the physical resonance states
does not exist in general.
It is worthwhile to mention that this mechanism had already been pointed out 
by Eden and Taylor more than four decades ago~\cite{eden-taylor}.
And thirdly, comparing the values between the bare masses and the Roper pole masses, 
one can see that the reaction dynamics can produce a sizable mass shift. 
It often comes to an issue that the Roper mass appears very high 
in the static hadron models.
However, in our point of view, it is not so surprising because the reaction dynamics 
are not taken into account in such static models.

\begin{figure}[t]
\begin{center}
\includegraphics[width=0.67\textwidth,clip]{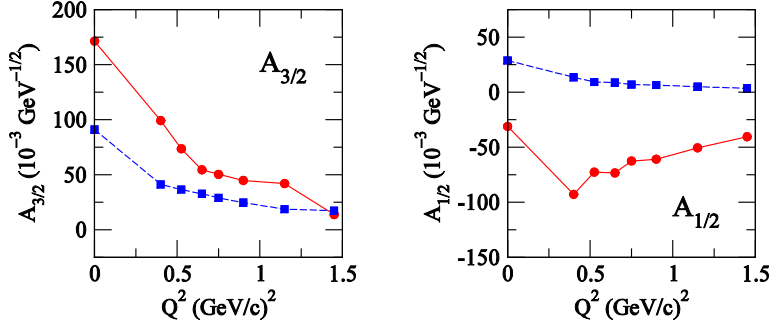}
\caption{$Q^2$ dependence of the $N$-$N^\ast$(1st $D_{13}$) 
electromagnetic transition form factors extracted with our early analysis~\cite{jklmss09,ssl10}.
Red circles (blue squares) with solid (dashed) line are 
real (imaginary) parts of the form factors.
\label{fig:d13-em}
}
\end{center}
\end{figure}

Figure~\ref{fig:d13-em} shows the electromagnetic transition form factors between the nucleon 
and first $D_{13}$ states, which were extracted by analyzing single pion electroproduction data 
from CLAS~\cite{clasweb}.
This is the first extraction of this form factors within 
a fully dynamical coupled-channels framework.
It is emphasized that the extracted form factors become complex, even though 
the corresponding bare form factors are purely real.
We observe that this complex nature of the resonance form factors 
is closely related with the fact that the $N^\ast$ states are not stable particles but resonances 
decaying to meson-baryon continuum states.

\section{Comprehensive analysis of $\pi N, \eta N, K\Lambda, K\Sigma$ production reactions}
\label{sec4}

In our early analysis as explained in the previous section, we took account of 
the $\gamma N$, $\pi N$, $\eta N$, $\pi\Delta$, $\sigma N$, $\rho N$ channels only, and
analyzed $\pi N\to \pi N$, $\pi N\to \eta N$ and $\gamma^{(\ast)} N\to \pi N$ reactions 
rather independently.
To make further progress, we have been extending our DCC model by including $K\Lambda$ 
and $K\Sigma$ reaction channels and performing the fully combined analysis of 
the $\pi N$, $\eta N$, $K\Lambda$, and $K\Sigma$ production reactions, in which 
all the model parameters associated with the hadronic and electromagnetic interactions 
are allowed to vary.
(See Table.~\ref{tab:history} for a comparison between our early and current analyses.)
This results in fitting to more than 20,000 data points.
The analysis is now almost completed and will be reported somewhere.

\begin{table}
\begin{center}
\begin{tabular}{lcc}
\hline
&2006 -- 2009 &2010 -- \\
\hline
 &6 channels& 8 channels \\
 & ($\pi N, \gamma N, \eta N, \pi\Delta, \sigma N, \rho N$)& 
($\pi N, \gamma N, \eta N, \pi\Delta, \sigma N, \rho N$, $K\Lambda$, $K\Sigma$) \\
\hline
$\pi N \to \pi N$            & $W < 2.0$ GeV~\cite{jlms07} & $W < 2.1$ GeV \\
$\gamma N \to \pi N$         & $W < 1.6$ GeV~\cite{jlmss08} & $W < 2.0$ GeV \\
$\pi N \to \eta N$         & $W < 2.0$ GeV~\cite{djlss08} & $W < 2.0$ GeV \\
$\gamma N \to \eta N$        &    ---        & $W < 2.0$ GeV \\
$\pi N \to K\Lambda$ &    ---        & $W < 2.1$ GeV \\
$\pi N \to K\Sigma$ &    ---        & $W < 2.1$ GeV \\
$\gamma N \to K \Lambda$     &    ---        & $W < 2.2$ GeV \\
$\gamma N \to K \Sigma$     &    ---        & $W < 2.2$ GeV \\
\hline
\end{tabular}
\caption{Summary for the dynamical coupled-channels analysis of meson production reactions at EBAC
in 2006-2009 (middle column) and since 2010 (right column).
\label{tab:history}
}
\end{center}
\end{table}

\section{Summary}
\label{sec5}

We have reviewed our 6-year effort of the nucleon resonance extraction via 
the dynamical couple-channels analysis of meson production reactions, which has
been pursued by a collaboration at EBAC of Jefferson Lab.
With this effort, we have successfully extracted the mass spectrum, decay widths, and 
various form factors associated with the nucleon resonances below $W = 2$ GeV.
Also, we have proposed an alternative view of the dynamical origins of $P_{11}$ nucleon resonances.
Our findings have revealed the critical role of the non-trivial multi-channel reaction dynamics,
which is usually neglected in the static hadron models,
in understanding the nucleon resonances.

\acknowledgments
The author would like to thank 
B.~Juli\'a-D\'iaz, T.-S.~H.~Lee, A.~Matsuyama, S.~X.~Nakamura, T.~Sato, and N.~Suzuki
for collaboration at EBAC.
The author also acknowledges the  support by the HPCI Strategic Program (Field 5
``The Origin of Matter and the Universe'') of Ministry of Education, Culture, Sports, Science
and Technology (MEXT) of Japan.
This research used resources of the National Energy Research Scientific Computing Center, 
which is supported by the Office of Science of the U.S. Department of Energy 
under Contract No. DE-AC02-05CH11231, and resources provided on ``Fusion,'' 
a 320-node computing cluster operated by the Laboratory Computing Resource Center 
at Argonne National Laboratory.

\end{document}